\documentclass[RNAAS]{aastex62}

\begin{document}

\title{Discovery of a White Dwarf-Main Sequence Binary in Open
  Cluster NGC 752}

\correspondingauthor{Andrew J. Buckner}
\email{abuckner@sdsu.edu, esandquist@sdsu.edu}

\author{Andrew J. Buckner}
\affiliation{San Diego State University, Department of Astronomy, San
  Diego, CA, 92182 USA}

\author[0000-0003-4070-4881]{Eric L. Sandquist}
\affiliation{San Diego State University, Department of Astronomy, San
  Diego, CA, 92182 USA}


\section{}

NGC 752 is one of the closest open clusters to the Sun, and it is old
enough \citep[about 1.5 Gyr;][]{twarog} that it should contain white
dwarfs (WDs), but to date none have been identified as members. Using
Gaia Data Release 2 \citep{GaiaDR2}, we searched a field of radius
$2\fdg5$ and selected members based on proper motions and
parallax. The combination of a proper motion cut and a parallax cut is
very effective at eliminating field stars from the sample (see top
panels of Figure 1). After correcting Gaia magnitudes for parallax and
extinction \citep[$E(B-V)=0.044$;][]{red752}, we identified an object
($\alpha_{2000} = 1^{\rm h}56^{\rm m}32\fs11$, $\delta_{2000} =
+38\degr19\arcmin02\farcs2$, $G = 18.785$, $G_{BP} - G_{RP} = 0.860$)
that fell nearly midway between the main sequence (MS) and WDs in the
color-absolute magnitude diagram (see lower right panel of Figure 1), and could plausibly
be a WD-MS binary. Although the measurements for such a faint object
have relatively large uncertainties ($\pi = 2.44\pm0.32$ mas,
$\mu_\alpha = 10.43\pm0.69$ mas yr$^{-1}$, $\mu_\delta =
-11.15\pm0.50$ mas yr$^{-1}$), it is the only WD object consistent
with cluster membership. It sits about $0\fdg56$ from the cluster
center, which is also within the cluster's halo on the sky.

\begin{figure}
\begin{center}
\includegraphics[scale=0.8]{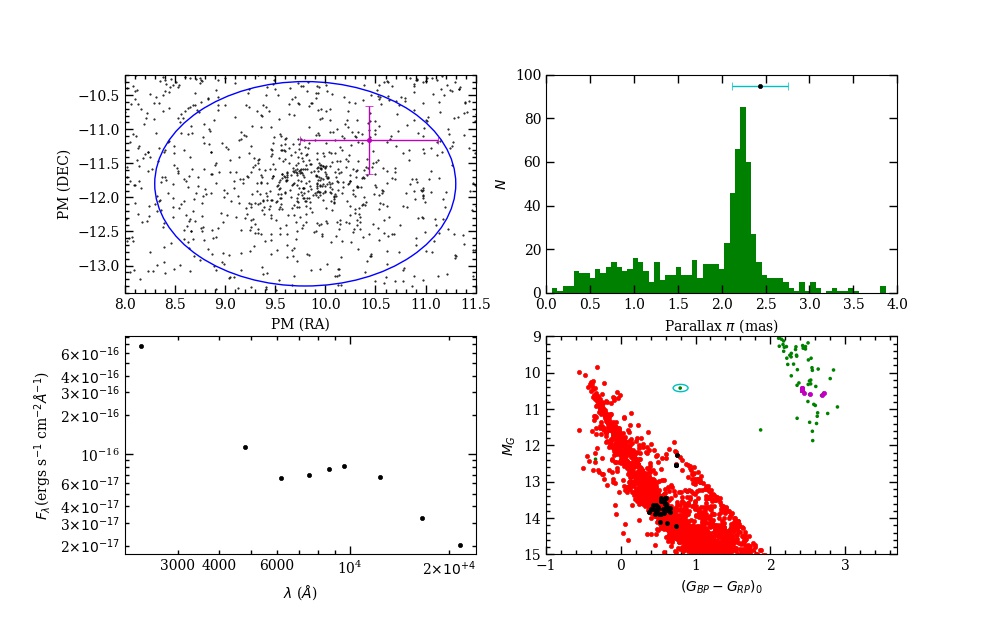}
\caption{Top left panel: Gaia proper motion cut applied to the NGC 752
  field, along with WD candidate values. Top right: Gaia parallaxes
  for proper motion-selected stars, along with the WD candidate value.
  Bottom left: SED from GALEX, Pan-STARRS, and 2MASS
  photometry. Bottom right: Color-absolute magnitude diagram for proper motion and parallax-selected NGC
  752 member stars ({\it green}), and nearby white dwarf stars
  ({\it red}) using Gaia DR2 parallaxes \citep{GaiaDR2}. Purple and
  black points are cluster and white dwarf stars whose combined light
  can reproduce the photometry of the suspected binary ({\it
    circle}).}
\end{center}
\end{figure}

To test this idea, we constructed a spectral energy distribution (SED)
from photometry in the ultraviolet from the GALEX \citep[$NUV$ filter;
][]{GALEX} survey, optical from Pan-STARRS \citep[$grizy$
  filters;][]{PS1}, and infrared from 2MASS \citep[$JHK_s$
  filters;][]{2mass}. Conversions from magnitudes to flux densities
were accomplished using the calibrations in \citet{GALEX} for GALEX,
\citet{Pan-STARR} for Pan-STARRS, and \citet{2MASS} for
2MASS. (Pan-STARRS and 2MASS use the AB magnitude system, while
GALEX uses a Vega magnitude system.)  The SED of the object shows
evidence of two peaks, indicating objects at very different
temperatures.

By looping through all combinations of faint Gaia WDs and NGC 752 MS
stars, we identified the combinations that could reproduce the
absolute magnitude $M_G$ and color $(G_{BP}-G_{RP})$ of the observed
object (see Figure 1, bottom right panel).  The comparison sample of
Gaia WDs was selected from stars with parallax $\pi > 20$ mas,
precision $\pi / \sigma_\pi \ge 20$, and a diagonal color-magnitude
cut $M_G > 10 + 2.6 (G_{BP} - G_{RP})$ \citep{GaiaHR}. The most likely
constituents are a MS star near the K/M spectral type boundary with
$M_G = 10.5$ and $(G_{BP}-G_{RP})_0 = 2.4$, and a WD with with $M_G =
13.6$ and $(G_{BP}-G_{RP})_0 = 0.6$. Based on the
spectroscopically-studied WDs found in this way, the WD probably has a
temperature near 6750 K.

The WD identified here is likely to be the brightest in the cluster if
further confirmed as a member.  For the future, photometric and
spectroscopic velocity monitoring could help clarify the nature of the
system, and whether or not there has been strong gravitational
interactions between the stars in the past. The object is outside of
fields with previous X-ray observations \citep{xray1,xray2}, and
future observations might detect it.

\facilities{Gaia, GALEX, PS1, FLWO:2MASS}

\end{document}